\documentclass[ aps,pra,showpacs,groupedaddress]{revtex4}
\usepackage{graphicx}
\usepackage{ulem}
\usepackage{latexsym}
\usepackage{color}
\begin{document}
\title { Real-time quadrature lock-in discrimination imaging through scattering media\\}

\author{Sriram Sudarsanam$^1$, James Mathew$^{1,2}$, Swapnesh Panigrahi$^3$,  Julien Fade$^3$, Mehdi Alouini$^3$,  and Hema Ramachandran$^1$.  
 }
\address {$^1$ Raman Research Institute, Sadashiv Nagar, Bangalore, INDIA 560080 \\ 
$^2$ Presently at University Hospital, Tubingen, D-72076 Tubingen, GERMANY)\\
$^3$Institut de Physique de Rennes, Universite de Rennes 1 CNRS, Campus de Beaulieu, 35042 Rennes, FRANCE}
\date{April 2015}
\maketitle

{\bf Abstract : Numerous everyday situations like 
navigation, medical imaging and 
rescue operations require 
viewing  through optically inhomogeneous media. 
This is a challenging task as  photons, instead of traversing 
ballistically,  propagate predominantly diffusively due to 
random multiple  scattering off the inhomogenieties. 
Imaging  {\it in real-time} with ballistic light  under 
continuous-wave  illumination  is even more challenging due 
to the  extremely weak signal, 
necessitating voluminous data-processing.  In this paper, we 
report imaging through strongly scattering media in real-time  
and at rates several times  the critical flicker frequency of the 
eye, so that motion is  
perceived as continuous.
Two factors contributed to the  speedup of {\it  more than 
three
orders of magnitude} over conventional techniques - the use 
of a simplified algorithm enabling
processing of data on the fly, and the utilisation of task and 
data parallelization capabilities of typical desktop computers.  
The extreme  simplicity and low cost promises  great utility 
of this technique. As an example, navigation under poor 
visibility is examined.
}

The difficulty of a ship to view a lighthouse source  due to fog, and the inability to 
directly see a bullet lodged in flesh are examples where recurrent random  
scattering in an inhomogenous medium degrades  the  image-bearing capability of 
light.  The normally ballistic transport of photons is rendered diffusive, precluding 
direct viewing of the source and inhibiting  formation of images (or 
shadows) of intervening objects. The need for imaging through 
scattering media, and to do so in real-time, perhaps felt since the dawn 
of mankind, continues to spawn fresh research
even today \cite{ Kang, Katz, Newman, Bertolotti}, as it has proven difficult to obtain a technique that is  simple, fast, compact and portable while also being versatile and inexpensive. \\
\indent The various  approaches pursued over the years 
\cite{CurrSc, Rudolph, French} either 
extract the minute amount of ballistic (forward scattered) 
photons from the overwhelmingly large amount of diffusive photons, or examine the diffusive 
photons themselves.  
Intuitively obvious is the  technique of time-gated ballistic imaging  
where an ultrashort pulse of light ($\sim$100 fs) illuminates 
the sample and the varying transit times of photons enables selection of ballistic light.
However, a far  
simpler and inexpensive approach is one where 
a continuous-wave source is modulated in intensity or 
polarisation  
and ballistic photons in the emergent light are identified based on their retention 
of the periodic modulation as 
opposed to the uniform temporal strength of the diffusive photons and ambient light. The 
exceedingly small proportion of ballistic photons
necessitates the acquisition of emergent light over a certain length of time, and 
a Fourier transform of the time-series enables their extraction 
to form two-dimensional images, as was first demonstrated  in 
Ref. \cite{OC}. This method of   
source modulation and Fourier-transform-based discrimination has now been 
extended to numerous modalities 
\cite { FDOCT,  struct-illum, GSA, SFD, Cuccia-Tromberg, Lee-Nam, fluo-nano, fourier-transform}.
However, despite availability of  well-optimized 
codes for Fourier analysis, 
real-time imaging, even for moderate scattering, remains elusive as
the large number of pixels in an image and the length
of the time-series  required to be examined render the computation
 voluminous. \\
\indent  A solution to this long-standing problem  is provided here by use of  a 
simplified algorithm - the quadrature lock-in discrimination (QLD) - for ballistic 
light  extraction,  coupled with efficient data-routing and hybrid parallelisation of 
tasks, that enable simultaneous  data acquisition and processing, leading to  
real-time display of images  with low latency and  at rates faster 
than the eye can perceive.  The utility and versatility of the technique is demonstrated 
in table-top experiments  that
simulate scenarios commonly encountered  in navigation. 
A scene containing several light sources 
is viewed through a scattering medium simulating the view from an aircraft   
approaching  a city airport on a foggy day. 
{The utility of modulated light 
sources and the efficacy of QLD 
in the elimination of clutter for  the  unambiguous identification 
of  runway  lights in the presence of other distracting sources is demonstrated, on the one 
hand, and in
the  easy and rapid visualisation of the 
runway under poor visibility on the other}. 
Thereafter, a passive scene is illuminated by modulated 
light, and the technique used to see both stationary and moving 
objects hidden from view,  affirming  
practical utility  of the technique in real-life situations. 

\section {Quadrature Lock-in Discrimination} 
\begin{figure*}
\vspace{-0.4cm}
\centering
\includegraphics[height=11cm, width = 18cm]{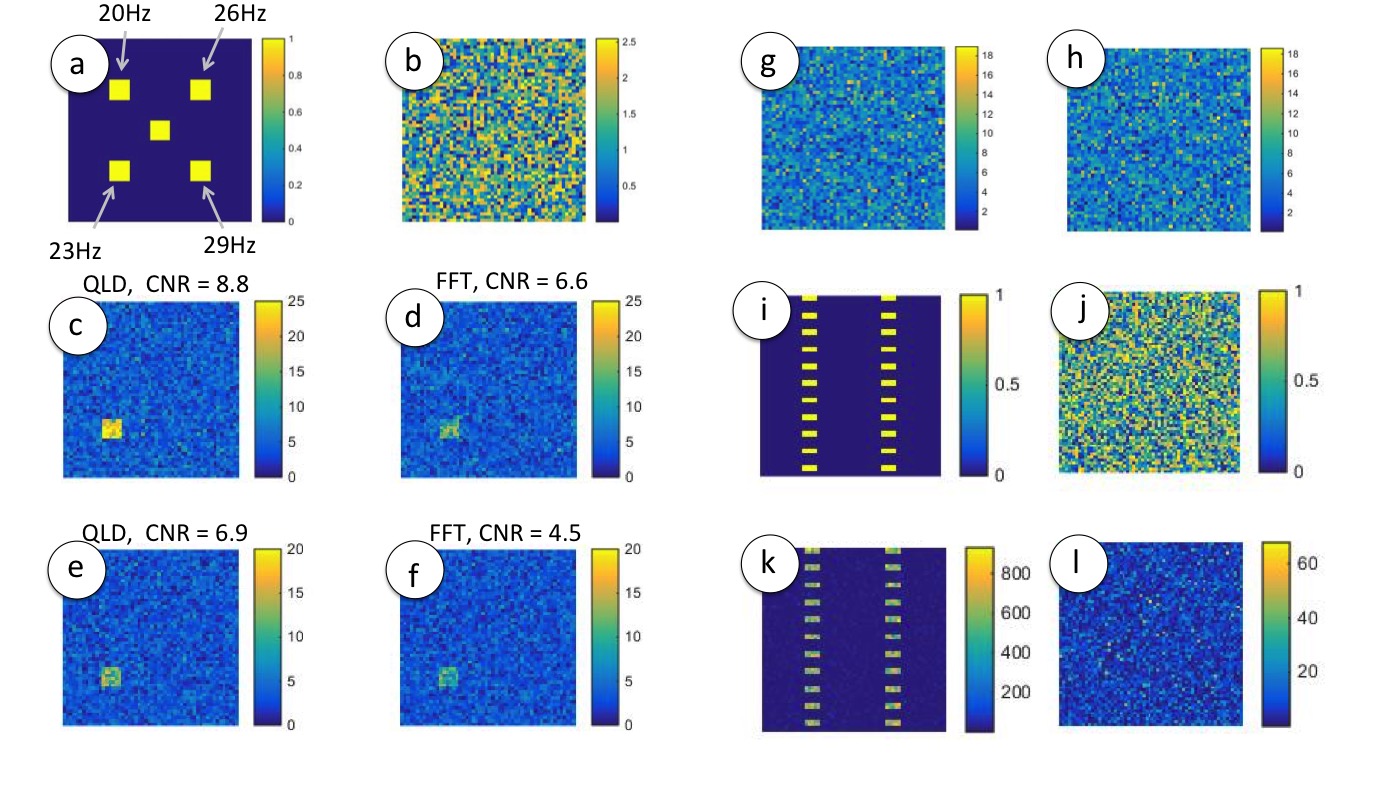}
\caption{\it{
Imaging through scattering medium by  use of QLD and FFT techniques  in a numerical example, 
and a suggested use of QLD for navigation. (a) A frame containing  five 
sources of light (intensities of four of which  are modulated sinusoidally while one 
is held fixed),  depicted at an instant when all sources are at their 
maximum intensity. In this example, the frequencies  of modulation   are 20, 23, 26, 
and 29Hz, and the frequency of interest is 23Hz. (b) A representative "raw data" 
frame where the random noise has been added to each pixel in the frames of the 
type (a), simulating a  frame  recorded on the camera  when the sources are viewed
through fog.  (c) Processed image,  obtained using QLD,   of the time series formed from N 
(82) frames of the type (b), displays a superior contrast-to-noise ratio in comparison to (d) which is obtained using FFT on the 
same N frames as in (c).  (e) Processed image with the source of interest  clearly seen,  obtained by performing QLD on N = 41 frames 
of the type (b). (f) Processed image obtained by FFT, using the same N = 41 frames as in (e).    
(g,h) Images obtained using QLD and FFT techniques, at an "incorrect" frequency, 31Hz, for the same data used in (c) and (d). 
QLD can be used, as shown in this computer simulated example, to detect modulated 
runway lights (i), which cannot be directly viewed under foggy situation (j). QLD at the 
correct frequency enables picking out the lights (k), which do not show up when QLD is 
performed at a different frequency (l). Details of simulation are given in Supplementary 
Information.}}
\label{fig:fft-qld}
\end{figure*}
\indent When the frequency of  modulation of the source is unknown, a fast Fourier 
transform (FFT) on the time series  provides  the amplitudes 
at various frequencies enabling the selection of the  dominant one.  
{{\it A priori} knowledge of the modulation frequency, 
however, 
eliminates the need for spectral decomposition -  a  fact utilised by  electronic lock-in detectors to extract weak 
signals from noise.  Time integration of the product of  
the recorded time-series  with an appropriately phase shifted 
reference sinusoid yields 
an output  proportional to the strength of the component at 
the reference frequency present in the 
original data. However, obtaining a two-dimensional image 
using a photo-detector and electronic lock-in detector 
requires a  step-scan over the array of pixels, as  
 in Ref. \cite{Fabien}.  The same idea may be implemented 
in 
software over an entire image, but  the computational complexity remains as the  
 phase difference between the modulating and the 
reference  sinusoids has to be determined.  Multiplying two 
sinusoids of the same frequency produces an output 
proportional to the cosine of the phase difference between 
them, and thus phase matching is desirable. The technique of 
Quadrature Lock-in Detection (QLD) circumvents the 
problem of the  actual phase determination  very simply,  by 
making a copy of the time series, and multiplying one by the 
reference sinusoid, and the other by a sinusoid $90^o$ phase 
shifted to this, so as to 
obtain the in-phase and quadrature-shifted components at the 
frequency of interest, both of which are then used to 
determine the amplitude at the frequency of
interest. \\
\indent We  first compare the performance  of QLD and FFT   
through numerical simulation of intensity-modulated light sources
being  viewed through fog.  In the absence of a scattering medium, the imaging camera would 
acquire, over the distinct regions corresponding to the images of the 
light sources, intensities varying sinusoidally at the respective frequencies (Fig.\ref{fig:fft-qld}a). 
In the 
presence of a scattering medium, the photons follow diffusive trajectories and are 
grossly deviated from their original paths; ballistic photons are significantly 
reduced in 
number. Every pixel in the camera now receives diffusive light. The 
regions corresponding to  the direct images of the sources too, 
receive predominantly diffusive light, along with a very small amount of ballistic light, 
the intensity of which  decreases exponentially with the strength of scattering. 
Consequently, each recorded frame shows diffuse illumination, with no  discernible 
feature (Fig.\ref{fig:fft-qld}b). A sequence of such "raw data" frames were 
generated numerically and QLD and FFT were performed on them (details in Methods).
While the five sources 
are hidden from direct view in any typical raw data frame,  the  appropriate source 
is revealed  when QLD   (or FFT) examines one of the modulating frequencies
(Fig. \ref{fig:fft-qld} c-f). 
However, when the  examined frequency does not match any of the modulating 
frequencies, no source is visible ( Fig.\ref{fig:fft-qld} g,h).  
It is seen that QLD offers a better noise 
rejection, yielding a higher 
contrast-to-noise ratio (CNR, see Supplementary 
Information) in comparison to FFT, for the same input data.  
For example,  QLD on a time-
series of 82  frames yields a CNR of 8.8, while FFT 6.6.  In 
fact, QLD on N/2 frames  yields a CNR comparable to FFT 
on N frames. \\
\indent This simple yet  effective technique may be put to use, 
for example, in navigation. Let us consider an airfield where 
the  
runway lights are  modulated at frequency $\omega_o$  (Fig.\ref{fig:fft-qld}i), that in the presence of fog, 
become  obscured from view, as depicted in   Fig.\ref{fig:fft-qld}j. A 
time series of such raw-data  frames is acquired and QLD is performed over them to extract 
the ballistic information. The runway lights 
are reconstructed (with a slight 
 noise) when QLD is performed at the correct frequency (Fig.\ref{fig:fft-qld}k); 
 these lights do not show up in the  processed image when QLD is performed at the incorrect frequency (Fig. \ref{fig:fft-qld}l). \\
\indent  A major advantage of such standard QLD approach 
is its immunity to the phase difference between the  modulation at the source 
and the  observer \cite{Dorrington, Japan}. The need for 
frequent phase determination that becomes necessary when 
the modulation has phase jumps, or in the presence of 
relative motion between source and observer, is now 
eliminated, making lock-in detection less cumbersome 
and more reliable. As phase search is no longer required,
QLD leads to a drastic reduction in computation, and hence 
faster image retrieval. The superior noise rejection of QLD, 
which directly translates to use of shorter time series, further
contributes to reduction in acquisition and processing time. 
In addition,  it doubles the bandwidth for the rate 
of modification of a scene (change in relative phase/ motion
between source and observer).  Another feature of great 
relevance is that the entire time-series has to be acquired for 
FFT to be 
performed. In contrast, processing may begin with the 
acquisition of the first frame, in the case of QLD - a fact 
that contributes significantly to the reduction of latencies. All 
these aspects are extremely important  in  applications where 
speed is of essence, as in navigation. Practical utility demands 
that  images be produced with near zero latency, that is, with 
negligible delay between the acquisition of data and the 
rendering of the image. 
Images must also be produced at rates faster than 30 frames per second (fps) to 
mimic natural vision with continuity of motion, and at twice that rate to avoid flicker (perception of variation in intensity) \cite{flicker}. Here we report implementation of QLD in software, performed at the 
modulation frequency, which provides the advantage of versatility, low cost and 
upgradability, while providing processed images in real-time with 
latencies of the order of milliseconds,  and frame rates of 100fps, limited only by the performance of the camera.\\

\begin{figure*}[h!]
\centering
\includegraphics[height=10cm, width = 17cm]{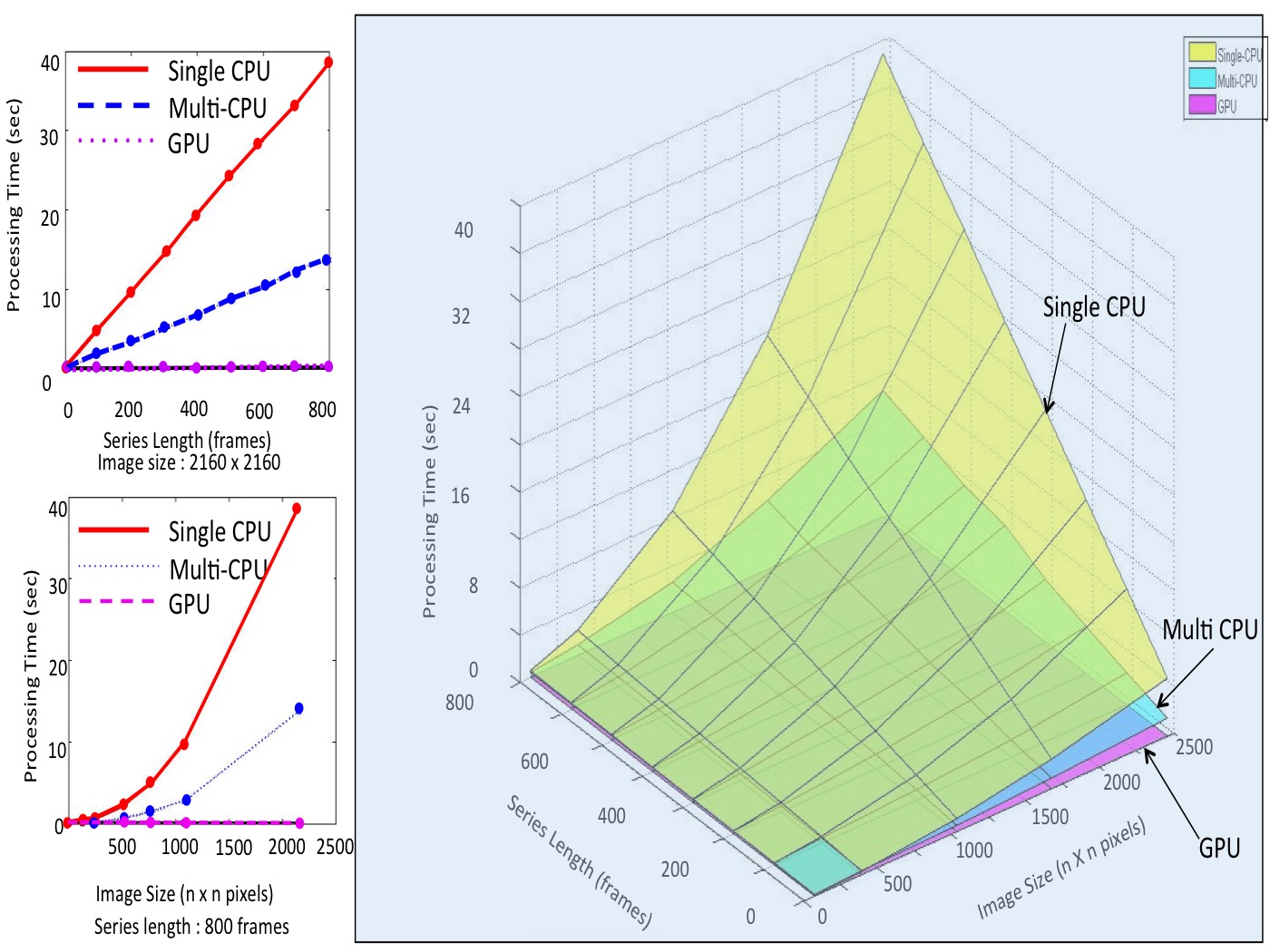}
\caption{\it{Latencies  for QLD when computed using a 
single CPU (yellow), multi-CPU of 4 cores with hyperthreading, 
resulting in a total of 8 threads (blue)  and a single GPU 
(pink).
}}
\label{fig-single-multi-gpu}
\end{figure*}

\section {Processing Time and Speedup}

\indent As the  Fourier 
decomposition  is essentially a series of 
QLDs being performed at different frequencies,   it is 
expected  that 
QLD at a {\it single} frequency can be executed faster.  However, several 
Fast Fourier Transform (FFT) algorithms 
have been developed that compute the Fourier transform not by the 
definition, but by  
methods of reducing the problem to smaller parts, so that the 
computation time 
is reduced from $O(N^2)$ to $O(N ln N)$,  N being the length of the time series. Heavily optimised 
commercial codes
are available that may be used for minimising  processing time, and thus, it may well be 
that a FFT consumes lesser time than QLD. 
We examined,
as function of the size of the frame and the length of the time-series, the time taken to obtain a processed image using 
FFT and QLD algorithms in MATLAB (see Supplementary Information). The time advantage of QLD 
over FFT was minor, implying that the FFT algorithm is indeed well optimised.   Next, a comparison was made of 
implementation of QLD using MATLAB and C++;   a speedup of nearly an order 
of magnitude was observed in the latter.  The processing time, nevertheless was 
still a few seconds. Clearly, such long processing times, with additional 
overheads of other tasks, cannot be tolerated if real-time images are sought. 
The phrase "real-time" allows for latencies of different durations in different 
contexts. For example, while imaging a static scene,  a time gap of $\sim10s$ may be tolerated. In medical applications, a 
latency of a few seconds qualifies as real-time. 
{ In fact, most often, "real-time" is taken to imply 
that  acquisition is not halted to enable processing 
and  data is being continuously  updated, albiet with a 
latency of several seconds (e.g. Ref. \cite{Boppart}). }
Navigation, however, imposes much more stringent demands on the speed of processing.
A boat approaching the coast, or a train moving at slow speed, would have moved
several 10's of meters in a few seconds. Thus, a delay of the order of $\sim$ seconds
in obtaining an image cannot be afforded; image latencies have to be reduced by at least two orders of magnitude. 
Further, as smooth movements require a display rate of 25fps or more,  and  
flicker-free viewing  60fps or more,  it is imperative that  the display of processed images and therefore the recording of 
raw data be  carried out at these rates or faster  and that the processing be completed within the 
time interval between acquisition of two frames, which is $\sim$15 milliseconds.  In the work presented here, this demand  is met  by parallelisation of tasks. Present-day desktop computers  boast of multicore 
Central Processing Units (CPUs)  and efficient Graphics Processing Units (GPUs), 
both of which allow for parallelization, but by different means.  Multicore CPUs enable 
simultaneous execution of different tasks running in different threads, working 
either individually, or in groups,  and thus 
are efficient for {\it task} parallelization.  On the other hand, GPUs, with their  
very large number of cores, facilitate simultaneous execution of the same 
operation on an array of  data - the so-called Single Instruction Multiple Data (SIMD) processing, offering {\it data } parallelisation.  The time taken for performing QLD on 
a given set of data using a single CPU, multi-CPU and a GPU are 
shown in Fig. \ref{fig-single-multi-gpu}. The use of 4-CPUs with a total of 8 threads  has reduced the processing 
time  from 38.7s to 14s for 800 frames of size 2160 x 2160.  The use of  a single GPU for computation brought the computation time  down to a mere 
20ms  -- a gain in speed  by  a factor of 700 over multi-CPU implementation and of {\it more than three orders of magnitude} over  the single-CPU implementation.\\
{\indent}	Thus far,  only  the computational  time with the 
respective algorithms has been considered. However,  the task  of obtaining a single 
processed image consists of the  acquisition of requisite number of  raw frames, 
transfer of data from the camera to the computer, the performing of  QLD by software, and the 
display and storage of the processed image.  In fact, it turns 
out that with GPU implementation of QLD,  computation  
consumes a very small  fraction of the entire time.  It now 
becomes imperative  
that  the time taken by the different non-computational 
tasks be considerably reduced. In this context, a very  important fact is that the C++ environment permits direct access to the camera control  (ANDOR Neo sCMOS camera, see Supplementary Information) facilitating rapid setting of camera's operating parameters, 
data routing, memory management and  display on the computer screen, all of which contribute to time 
overheads, though to varying extents, and  have to be carefully optimised. 
Leveraging the distinct  advantages of CPUs and GPU;  we employ hybrid 
parallelisation where  multiple 
CPU cores are employed  for different tasks
such as acquisition, buffer management and kernel calls and the GPU 
is used to perform  QLD  on the arrays of pixel data, invoking SIMD  (see Supplementary Information). With this strategy, we have been able to produce  processed images of 
objects originally obscured by scattering  {\it{\bf  in real-time, in a sustained manner}}.   Frames of size $600 \times 600$ pixels (3.6 
mega-pixels) are displayed at 100 fps, with 
the first processed image appearing 55ms after the acquisition of the first raw data 
frame, and 5ms after the acquisition of the last raw data frame of the sequence of 
images required for QLD. This near-zero latency and camera-limited frame rate, that can be maintained over long term operation,   
paves the way for real-time imaging in scenarios where speed is of essence, as in navigation. \\

\section {Experimental Demonstration} 
\begin{figure*}[h]
\centering
\includegraphics[height=14cm, width=18cm]{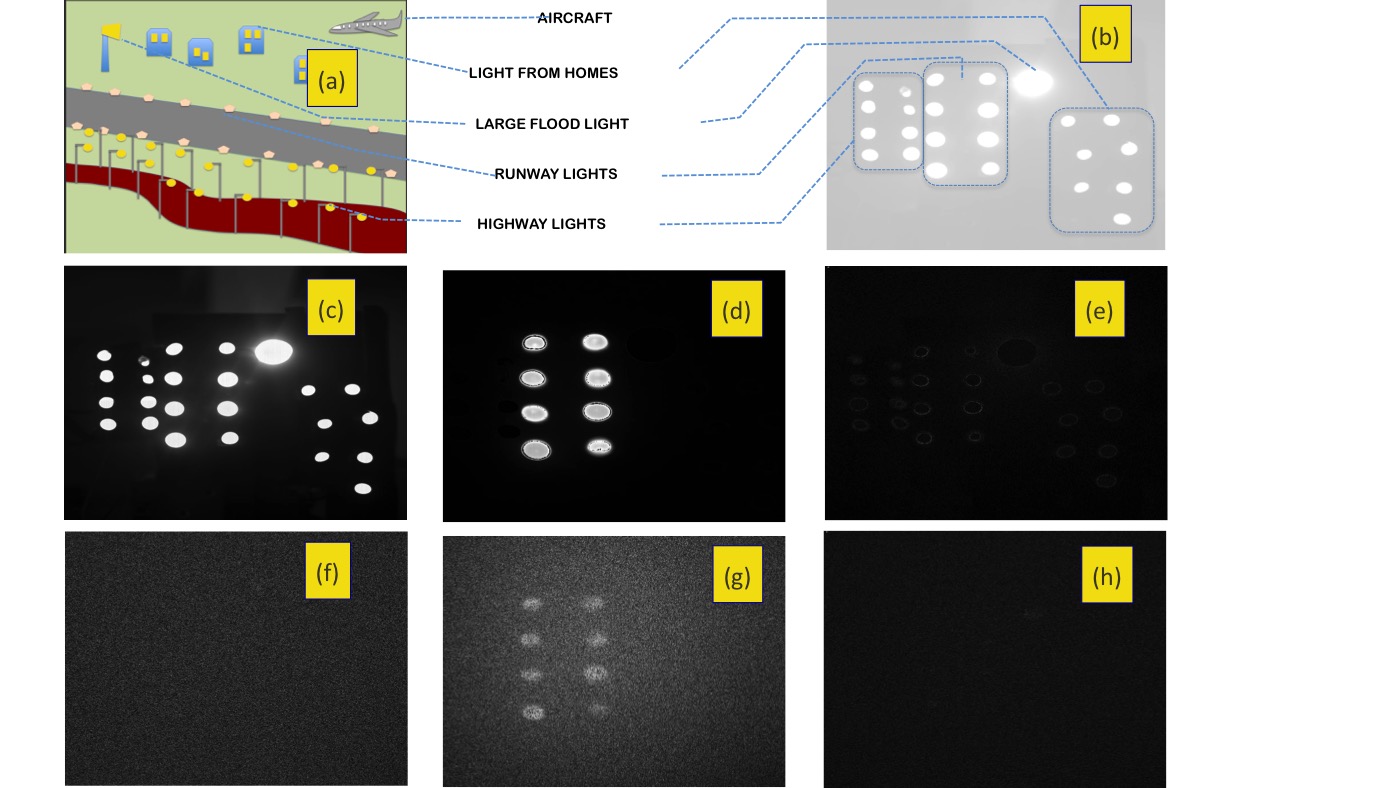}
\caption{\it{ Real-time imaging of light sources obscured by scattering medium:  (a,b) The scene being simulated experimentally is one that has a variety of 
light sources. (c) The scene as it  would appear to a pilot approaching for landing 
on a clear night. (d) Image obtained using QLD, where only the modulated runway lights are picked up. (e) No light source shows up when QLD is attempted at the 
incorrect frequency. (f) Typical view on a foggy night. (g) QLD at the correct frequency shows the runway lights. (h) QLD at the incorrect frequency shows no source.\\ }}
\label{fig-runway-expt}
\end{figure*}

\indent We now demonstrate such
real-time imaging in several scenarios, in tabletop 
experiments that mimic every-day situations.  Uncollimated, incoherent light emitting diodes (LEDs) 
were used as the sources 
of light. These were powered by dc sources, and the desired LEDs could be 
modulated in intensity at the required frequency. 
{ 
The use of QLD essentially requires a sinusoidal modulation of the source in 
some form, e.g., intensity modulation. However,  other considerations may require 
that the intensity remain uniform. In such cases,
one may modulate the  direction of linear  polarisation of the source  in a sinusoidal fashion and perform a 
polarisation based QLD on the emergent light  after  
passing it through a fixed analyser \cite{OC}.  Discrimination is 
based on the fact that the ballistic, or the forward scattered  light retains its 
polarisation, while the diffusive light and ambient light 
are depolarised to various extents.  The experiments 
discussed below, however,  employed intensity modulation. }The scene was viewed using  the ANDOR Neo sCMOS camera and the 
data acquisition and real-time processing was performed using the  parallel 
processing procedure described in Supplementary Information. 
 Frames of size $600 \times 600$ 
pixels, with exposure time 5 ms were recorded at 
100 fps. Shorter exposures could be  used when the scattering was less.   The first 
scenario 
mimics the pilot's view while approaching  a runway for landing. In addition to the 
runway lights, there is a  clutter of other sources of light - streetlights, 
lights in buildings,  vehicular lights, etc. (Fig. \ref{fig-runway-expt}a,b,c). 
Modulation of the 
runway lights, either in intensity  or in polarisation,  enables the  use of QLD to  
reject all 
light  except that  from the sources  of interest. These alone appear in the 
processed image
 when QLD is attempted 
at the modulation  frequency (Fig. \ref{fig-runway-expt}d), and are not visible at any other 
frequency (Fig. \ref{fig-runway-expt}e). 
Thus, QLD may be used, even when visibility is good, to identify particular sources of interest in the presence of a large number of unwanted sources in the field of view. 
Next,  we simulate a foggy day (or night) where  the  same set of sources (runway  lights and other sources) is obscured,   that is, the  
lights cannot be discerned   in  snapshots of the scene (Fig.\ref{fig-runway-expt}f). This was achieved by 
interspersing between the scene and the camera a glass container 
with a scattering medium, that has   
spherical polydisperse scatterers, ranging in size 
from   0.5$ \mu$m to 5 $\mu$m, typical of water droplets in 
atmospheric fog. The scattering medium simulated 262 m of moderate fog, or 26 m
of dense fog (see Supplementary information). QLD at the correct frequency reveals the 
light sources of interest (Fig. \ref{fig-runway-expt}g). No light source shows up when QLD is performed at a different 
frequency (Fig. \ref{fig-runway-expt}h).   Processed images of size $600 \times  600$ 
pixels were obtained at 100 fps, with a latency of  
5 milliseconds.  Thus, with regard to  
continuity of motion and immediacy of view, the processed images  appear to 
the pilot akin to normal  vision.\\

\begin{figure*}[h]
\centering
\includegraphics[height=12cm, width=16cm]{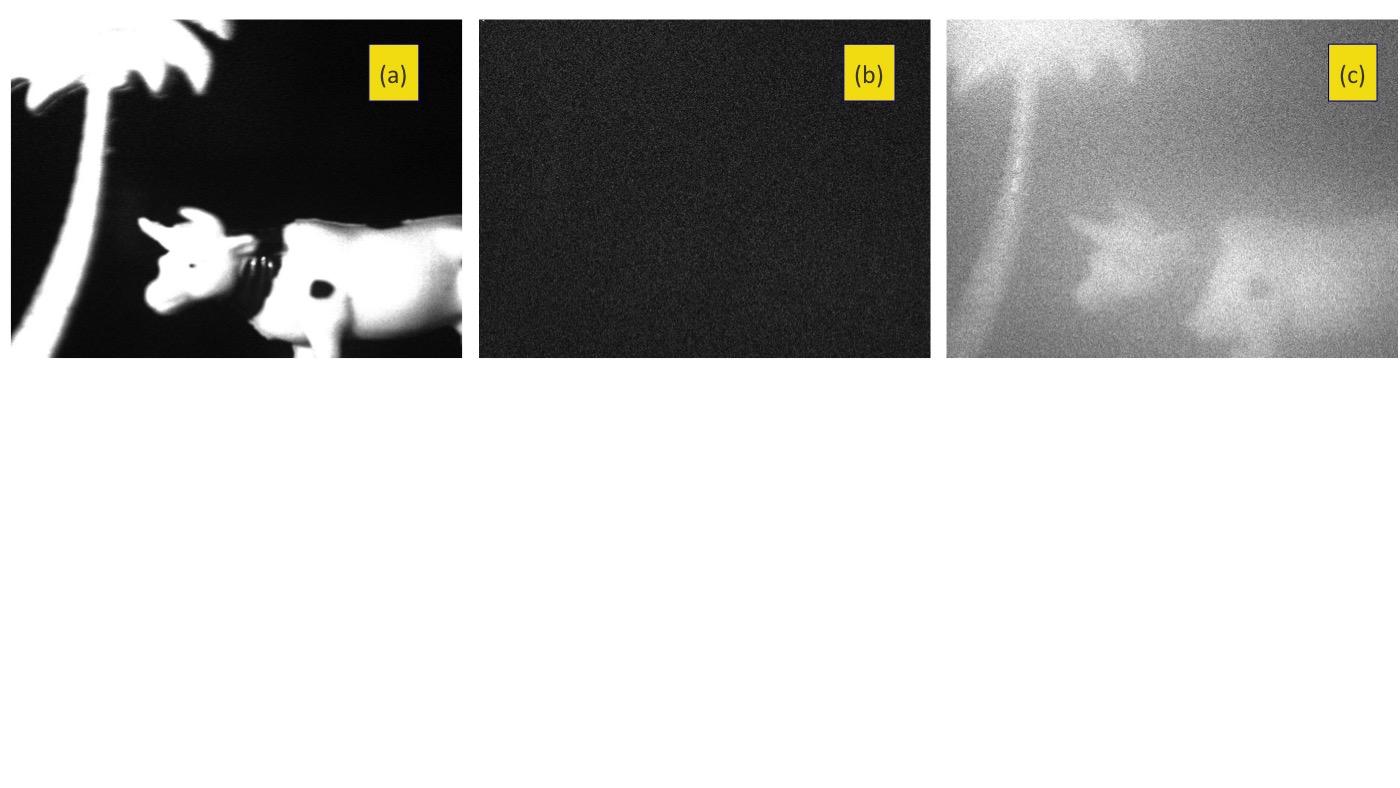}
\vspace{-6cm}
\caption{\it{ Real-time imaging through scattering medium 
: (a) Model of a tree and a cow, kept in the field of view. (b) A 
typical camera image when the  tree and the cow are 
obscured by a strong scattering medium interspersed between 
the scene and the observer. (c) Upon QLD, the tree and the 
cow become discernible. }}
\label{fig-cow-expt}
\end{figure*}
\indent  The second scenario  simulated  was one often 
encountered while driving in the countryside in fog.  Unknown terrain, a 
curve in the road, a tree in the path, a boulder fallen onto the road, cyclists, or 
even animals in the path cannot be made out from the raw images of the scene 
captured by a camera on the vehicle. If, however, the illumination (car fog-lights, say)
is modulated in intensity or polarisation, and realtime QLD performed on 
the captured images, these objects can easily be made out, as illustrated in Fig. 
\ref{fig-cow-expt}. Models of a tree and an animal were placed in the field of 
view of the 
camera (Fig. \ref{fig-cow-expt}a). 
Fog, simulated by the strongly scattering medium described earlier,
obscured these from view (Fig. \ref{fig-cow-expt}b). 
However, performance of QLD enables one to see these objects (Fig. 
\ref{fig-cow-expt}c). These images were also obtained in realtime at 100 fps, on processing data acquired over
 50 milliseconds. Sharper contrast can 
be attained by processing over a longer time-series. The same display frame-rate of 
100fps can be maintained; the first processed image would appear after a longer 
delay from the first acquired raw data frame, though the gap between the last acquired 
raw data frame and appearance of the first processed image would still be $\sim$5ms. 
\\
\begin{figure*}[h]
\centering
\includegraphics[height=9cm, width=16cm]{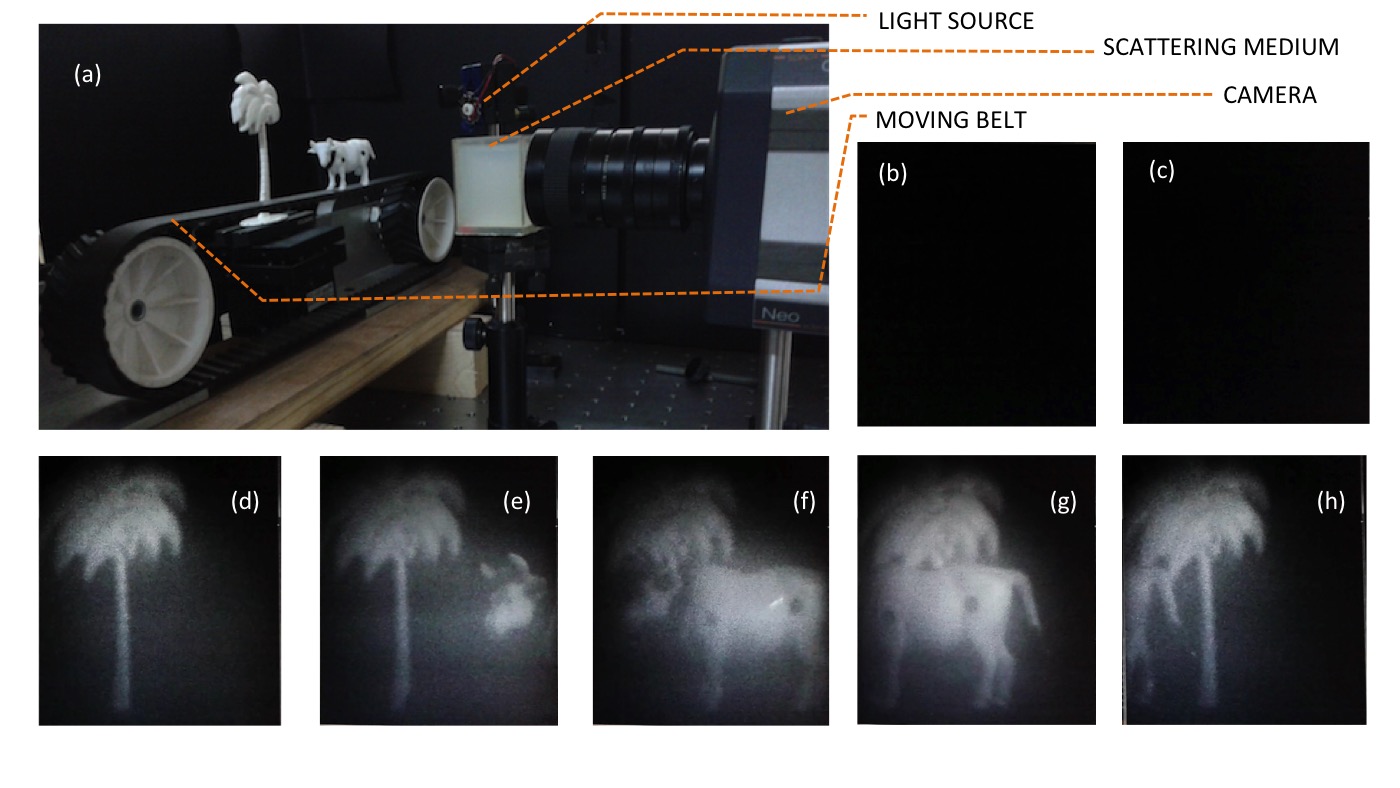}
\caption{\it{Realtime QLD imaging of a moving object (a) 
The experimental setup where models of a tree  and a cow 
illuminated by a modulated source,
though in the field of view of a camera, are obscured due to 
an intervening  scattering medium. The cow is on a moving 
belt so that it moves across the scene from right to left. 
(b) A typical raw data frame recorded by camera, (c) 
processed QLD image when performed at the "incorrect" 
frequency, (d-h) snapshots from a real-time movie displaying 
processed QLD images (at the "correct" frequency), showing 
the moving object clearly. Processed images of $600 \times 600$ pixels, are obtained at 100fps, are obtained 55ms after recording the first frame and 5ms after recording the last frame of the time series.   }}
\label{fig-moving-cow-expt}
\end{figure*}

{Finally, we demonstrate real-time imaging through scattering media using QLD, 
when there is a relative movement between the object and the viewer, as would be 
the case in a navigation.  The changing separation between the object and 
the observer adds to the complexity in modulation-based imaging as this 
continuously alters the relative phase between source and observer.  Thus, to 
obtained good images,  with minimal smearing,  it is important that  
inter-frame delays be low and that images be extracted from shorter time series. The 
second experiment was repeated, this time with the
model of the animal placed on a moving belt (Fig. \ref{fig-moving-cow-expt}a). Once again, the scene illuminated by 
modulated light was viewed  through a scattering medium. 
Figures \ref{fig-moving-cow-expt}b and 
\ref{fig-moving-cow-expt}c show a typical 
raw data frame and a processed image when QLD is 
performed at an "incorrect" frequency. In both cases, no 
object can be discerned. Figures \ref{fig-moving-cow-expt}d -
\ref{fig-moving-cow-expt}g  
are snap-shots from a movie taken of a computer 
screen that displays the processed images of QLD 
performed at the correct frequency. Both the stationary 
object (tree) and the moving object (cow) can be seen quite 
well. In this experiment, raw frames were acquired at 100fps, 
and processed images generated at the same rate. The  latency 
was 5ms, hence compatible with realtime flicker-free display to the human eye.}

{\indent} The three experiments show that  the technique is  capable of providing real-time, clutter-free images through scattering media.  While many modern airports have sophisticated instrument landing systems, the 
technique presented in this article is likely to find utility for various 
 other forms of navigation  - small aircraft in private fields, rail and road 
travel and  maritime travel. The technique  is equally well applicable to other  areas,
 like  imaging  through flesh, 
 rescue operations in fires, and deep ocean viewing,   that demand the mitigation of 
effects of multiple scattering and where speed is of essence.  Some situations of medical imaging, e.g., looking at 
a beating heart, 
impose an upper limit on the time afforded for data capture, while viewing moving 
objects like a victim in a smoke-filled room \cite{fire},  or saving a drowning person, also restricts the time 
afforded for processing. \\
\indent To summarise, we have used  modulated, continuous-wave  incoherent light 
sources and performed QLD on the light 
emerging from a strongly scattering medium to
discriminate the ballistic light from the diffusive.  This, in 
conjunction with  
hybrid parallelisation, has enabled, in a sustained manner,  the 
visualisation of obscured sources and hidden objects
in real-time, with millisecond latencies and at frame rates far 
exceeding  the usual refresh rate of TV movies (25 fps) and the 
critical flicker frequency of the human eye (60 fps).
 The versatility of the technique has been  demonstrated in 
three different 
scenarios commonly encountered in navigation.  By virtue of 
its simplicity, 
extremely low cost, and portability, the technique demonstrated 
here has 
enormous potential  for application, providing an  interesting
alternative to the well-established  time-gated ballistic imaging 
with pulsed light. 

\section {Acknowledgements}
We wish to thank B.S. Girish, M.S. Meena and B. Ramesh for several discussions.
We thank the Indo French Centre for Promotion of Advanced Research, New Delhi, for funding of the project, and Department of Science and Technology, India for the laboratory facilities.

\section {Methods}
\noindent {\bf Numerical simulations : }
The scene being simulated is one that has five light sources, four of which 
are sinusoidally modulated in intensity while the fifth is held constant. 
A typical camera frame capturing the scene  is simulated by 
a frame of $n \times n$ pixels, that has within it,  
five non-overlapping  sub-regions of size $ m \times m$ pixels, corresponding to 
the direct image of the  sources (Fig. \ref{fig:fft-qld}a). A number (N) of such 
frames are created 
to represent subsequent snapshots separated by  short time intervals ($\Delta$t); the intensities 
in four of the  sub-regions are varied sinusoidally as 
$(0.5 + sin[\omega_i t_s]),  i = 1,..4$ and $t_s = s \Delta t $,  where $s = 1, 2, ...N$.  The 
arrival of diffusive photons is simulated by adding to every pixel in each frame a 
(different) random number, uniformly distributed between 0 and X, where X 
depends on the 
strength of scattering (simulations were carried out for X ranging from 1 to 10). 
This results in  a series of noisy frames like the one in  Fig. \ref{fig:fft-qld}b, where the sources 
cannot be directly seen, even though information about  the modulated 
sources is contained in it. From the  sequence 
of  $N$ such frames, time series are generated for each of the $n^2$ pixels, by 
selecting the intensity values for that  pixel from successive frames recorded at 
time instants $t_s$, i.e., an array $I(j, k, t_s), s = 1, 2, ...N$ was formed for every pixel 
(j, k) of a frame. The contents of these $n^2$ time series simulate  intensity 
information of light  due to  spurious 
unmodulated ambient illumination, diffusive  light from the source,  a minute  
additional sinusoidal contribution simulating the arrival of ballistic photons for pixels 
in the sub-regions $i$, 
and also electronic noise. The aim is to 
extract 
the ballistic component from the source of interest.\\ 
\indent Using a Dell Precision T-3600 desktop computer  and programming in 
MATLAB,  we have compared the 
performance of FFT and QLD  techniques of extraction of the ballistic photons. 
For the former, the FFT  function was used, and $\mathcal{F}(j,k,\omega_q), (q=  -(N-1)/2, ..0, 1, ...N/2)$, the Fourier transform of $I(j,k,t_p)$ (with p = 1, 2, ....... N) was evaluated. The  map of the 
$ |\mathcal{F}(j,k,\omega_{q=i)}|^2$  gave the image with the ballistic component that retained 
modulation at frequency $\omega_i$ (Fig. \ref{fig:fft-qld}(c)). In the case of QLD, the quantity 
 $R(j, k) = 
[{\Sigma_s[I(j, k, t_s) . sin (\omega_i t_s)]}^2 + {\Sigma_s[I(j, k, t_s) . cos (\omega_i t_s)]}^2]^{1/2}$ 
 was evaluated for each pixel, (j, k), with the values of $sin (\omega_i t_s)$  
and $cos (\omega_i t_s)$ being read from pre-calculated arrays. The 2-d plot 
of R(j, k) gives the image due to the ballistic photons (Fig. \ref{fig:fft-qld}(d)). 
Both calculations were performed for the same set of recorded frames, and for 
the same length of time series. \\
\noindent {\bf Estimation of Contrast-to-noise-ratio : }
We define \cite{Ritfold} contrast-to-noise ratio as 
\begin{equation}
CNR = \frac {\overline{I}_{s}\, -\, \overline{I}_{surr} }{\big{[}\,\,\overline{(I(j,k) \,-\, \overline{I}_{surr})^2}\, \,\big{]}^{1/2}}, 
\end{equation}
where 's' denotes the source and its intensity $I_s$ is   averaged over the $m \times m$ 
pixels corresponding to the source of interest;  'surr' denotes a square slice surrounding 
the source of interest and its average is 
perfomed over the $8m^2$ pixels falling in the $m$-pixel wide square slice (see Fig. \ref{surround}). 
The sum and average in the denominator is, likewise,  performed over these $8m^2$ 
pixels. The CNR thus provides a measure of how prominent the source is over the 
surrounding region, in units of the standard deviation of intensity in the surrounding 
region.  \\ 
\begin{figure}
\includegraphics[height=6cm, width=12cm]{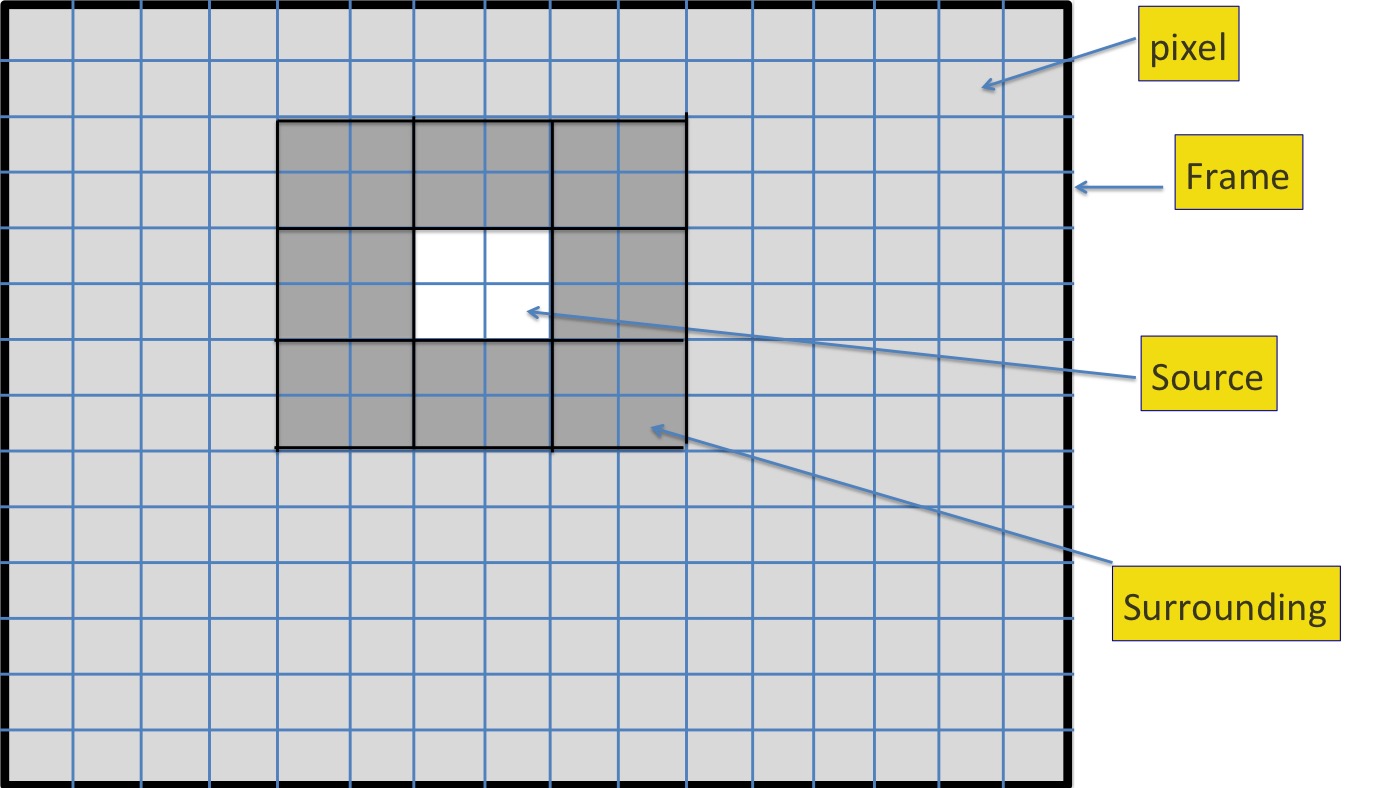}
\caption{ Example of a frame with 14 x 16 pixels, with a  source (white region)  of size 4x4 pixels. The 
surrounding region considered for determining CNR  is made of 8 such 4x4 squares, shaded dark grey.}
\label{surround}
\end{figure}

\noindent {\bf Estimation of depth of fog  : }
According to Ref. \cite{Navy} moderate fog has a visibility of 125m; i.e., light flux reduces to 1/e of its initial value 
over this distance. The scattering medium used had an optical depth of 2.1 
and is thus equivalent to 262m of moderate fog or to 26m of dense fog. 
According to current  aviation rules regarding Instrument Landing Systems (ILS) \cite {DGCA}  a visibility of 300m is required for  catergory-II landing, 
and of 175m for category-IIIA landing. The experiments thus simulate situations where even
category-III landing cannot be permitted in case of moderate fog.  We show that using QLD and parallel 
processing, real-time images can be  obtained over 262m of moderate fog, thus increasing visibility. This is 
illustrated in  Fig.\ref{fig-runway-expt}g, 
where on performing QLD the runway lights  come into view.  Once again, QLD at the incorrect 
frequency yields no image (Fig. \ref{fig-runway-expt}h). \\

\section {Supplementary Information}

\noindent{\bf Details of apparatus :}
We used an ANDOR Neo  sCMOS camera that has  2560 x 2160 pixels and is  capable
of  acquiring, in a sustained mode,  full frames at 20fps, and  smaller frames of 128 x 
128 pixels  at 1500fps.  Data is transferred to the computer  over a  Camera Link cable.
The camera was controlled by our C++ program, using commands provided in  
ANDOR's software development kit.  The camera was maintained at room temperature. Intensity data was recorded with 16-bit resolution.  

The computation was carried out on a  Dell Precision T3600 desktop computer that has a  Xeon Quad Core  processor running at 
3.4 GHz, equipped with  64GB RAM. Hyperthreading  has been enabled in the four CPUs, so that a total of 8 threads can run simultaneously. However, at a time only six were used, as the computer 
was found to overheat beyond that.   The GPU of the desktop 
was QuadroPro 2000 running at  800 MHz and had  96 CUDA cores.  No 
specialised hardware was used. In fact, the same GPU was also used by the 
operating system for display. Likewise, all parallelisation functions used were  from 
the OpenSource library, e.g.,  Open-Multicore-Processing (OpenMP) for data 
parallelization in CPU, Portable Operating System Interface - POSIX (pThreads) 
library for task parallelization in CPU and CUDA for utilizing the computational 
power of the  GPU.  All computations were performed in double precision. 

\noindent{\bf Comparison of computational times for QLD and FFT  :}
Using a single CPU and using MATLAB (without parallelisation), a comparison was made of the 
computational times for performing the QLD and FFT operations 
for different frame sizes and lengths of time series. The results are shown in 
Fig. \ref{fig-matlab}. QLD is found to have a slight advantage over FFT, for the same input data. This slight advantage, however, acquires significance when
we recognise the fact  that to produce images with  a
given contrast-to-noise ratio, shorter time series suffice for  QLD  when compared to FFT. 

\begin{figure*}[h]
\centering
\includegraphics[height=10cm, width=18cm]{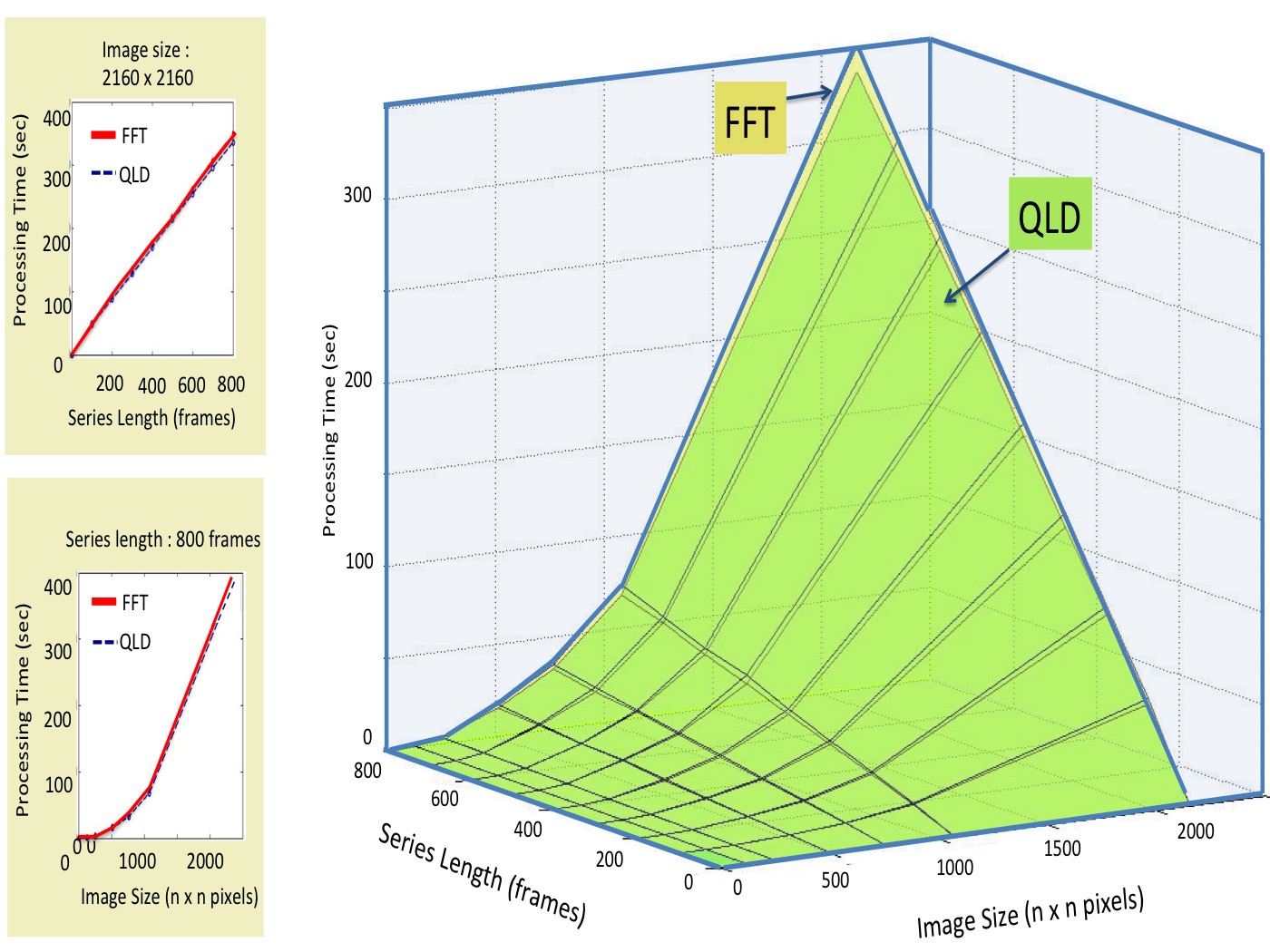}
\caption{\it{Comparison of the processing times for QLD and FFT, using MATLAB for computation.} }
\label{fig-matlab}
\end{figure*}

When using C++ instead of MATLAB for performing QLD, considerable speedup was observed. 
Processing times dropped down from  340s to 45s for  800 full-frames (Fig. \ref{fig-matlab-C}). 

\begin{figure*}[h!]
\vspace{-0cm}
\centering
\includegraphics[height=10cm, width = 17cm]{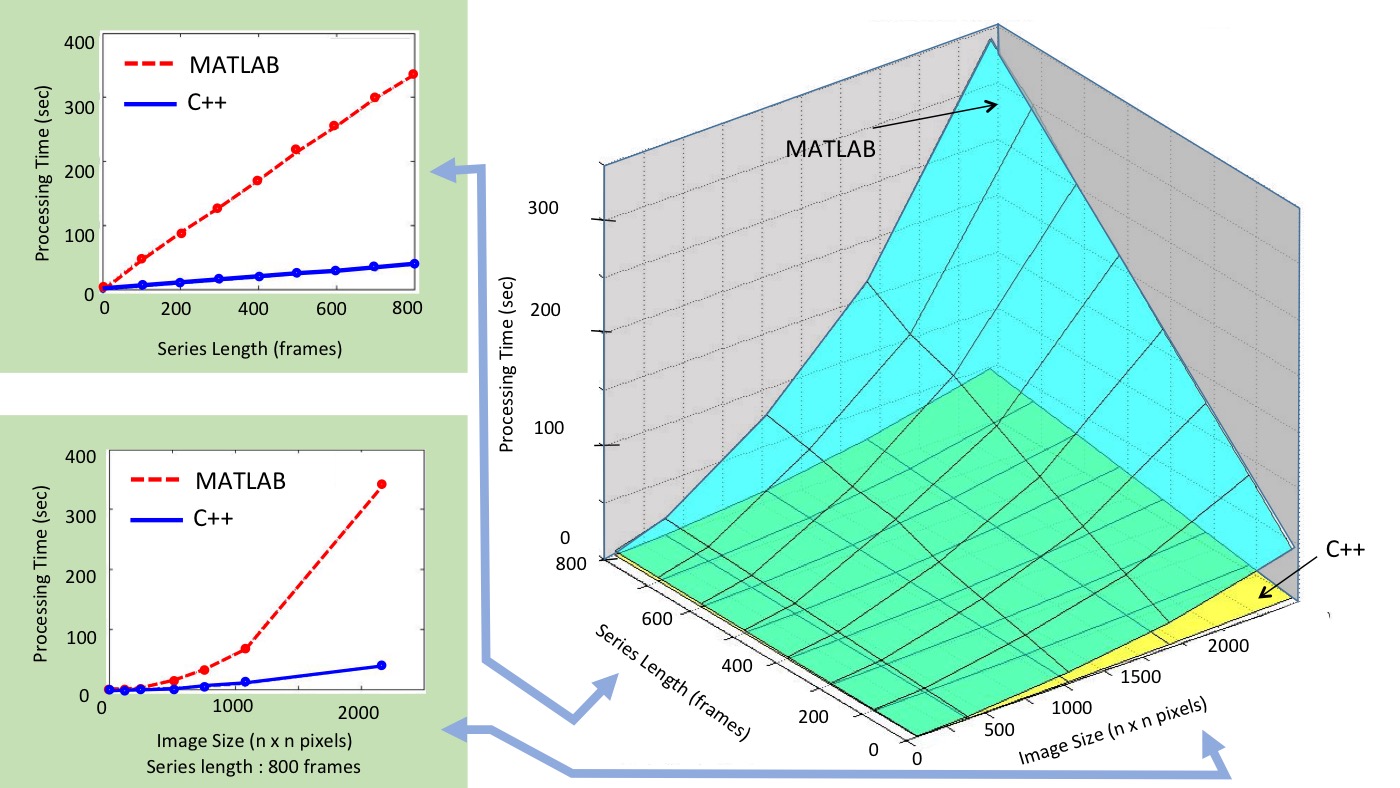}
\caption{ \it{Comparison of the processing times for QLD  using MATLAB and C++ for computation.} }
\label{fig-matlab-C}
\end{figure*}

\noindent{\bf Timing sequence for hybrid parallelisation :}
While the camera has its own software for setting the parameters and acquiring 
data, considerable time benefit was 
achieved by writing our own C++ codes. 
To obtain real-time images, raw-data  frames must  be processed as they arrive to 
the system from the camera. To facilitate this, the stream of data that the camera 
sends must also be stitched and arranged, facilitating processing.  The following 
is the sequence adopted. (See Fig.  \ref{fig-timing-sequence})
\begin{itemize}
\item  The program begins with the main thread (T1) declaring the necessary variables and 
setting the physical parameters specified by the user. These include the temperature 
of the camera, exposure time, and the sampling frequency. The thread then allocates the 
memory space in the Random Access Memory (RAM) in accordance with  the parameters chosen and creates tables of  
sine and cosine functions, appropriate for the choice of lock-in and sampling 
frequencies. Thereafter acquisition is initiated. A stream of 
16-bit binary data (2-bytes) now 
emerges from the camera and is carried over the Camera-link cable to the 
computer. 
\item {As bytes of data  stream in, thread T2 accesses the 
buffer to stitch pairs of bytes and store values in an array of the appropriate data type 
to  form the raw frame matrices.  Upon completion of the 
task, a flag is set to indicate that the data is in a form ready 
for analysis. }
\item { Thread T3   invokes  the Complete Unified Device Architecture (CUDA) kernel, which prompts the 
compiler to execute a certain  block of program  in the cores of the GPU.   As the 
GPU does not have direct access to RAM, thread T3  fetches data (raw frames); this transfer takes place over the PCI express bus at about 100Mbps.  
Intensity data at each pixel is multiplied by sin and cos 
to obtain two intermediate frames.   Successive  intermediate 
frames are accumulated into two different 2-D arrays that represent the two quadrature 
components. 
After  accumulation of the required number of terms, the components  are squared and 
added, resulting in the 32-bit  $n \times n$ processed image. This  is then  displayed on the monitor  using Open Source Computer Vision (OpenCV).}
\item{Processed data may be stored  to the harddisk, if required.} 
\item{ Flags, indicative of the 
current status of the threads, are set  to ensure proper communication between the 
threads and to avoid
race conditions.} 
\item{Although it appears that only three threads are running, each of the threads spawn further threads to facilitate data parallelization. For example, four hardware threads arrange the bytes and arrange the raw data, while the fifth communicates with the graphics card. The sixth hardware thread runs the main sequence. }
\end{itemize}

\begin{figure*}[h]
\centering
\includegraphics[height=10cm, width = 18cm]{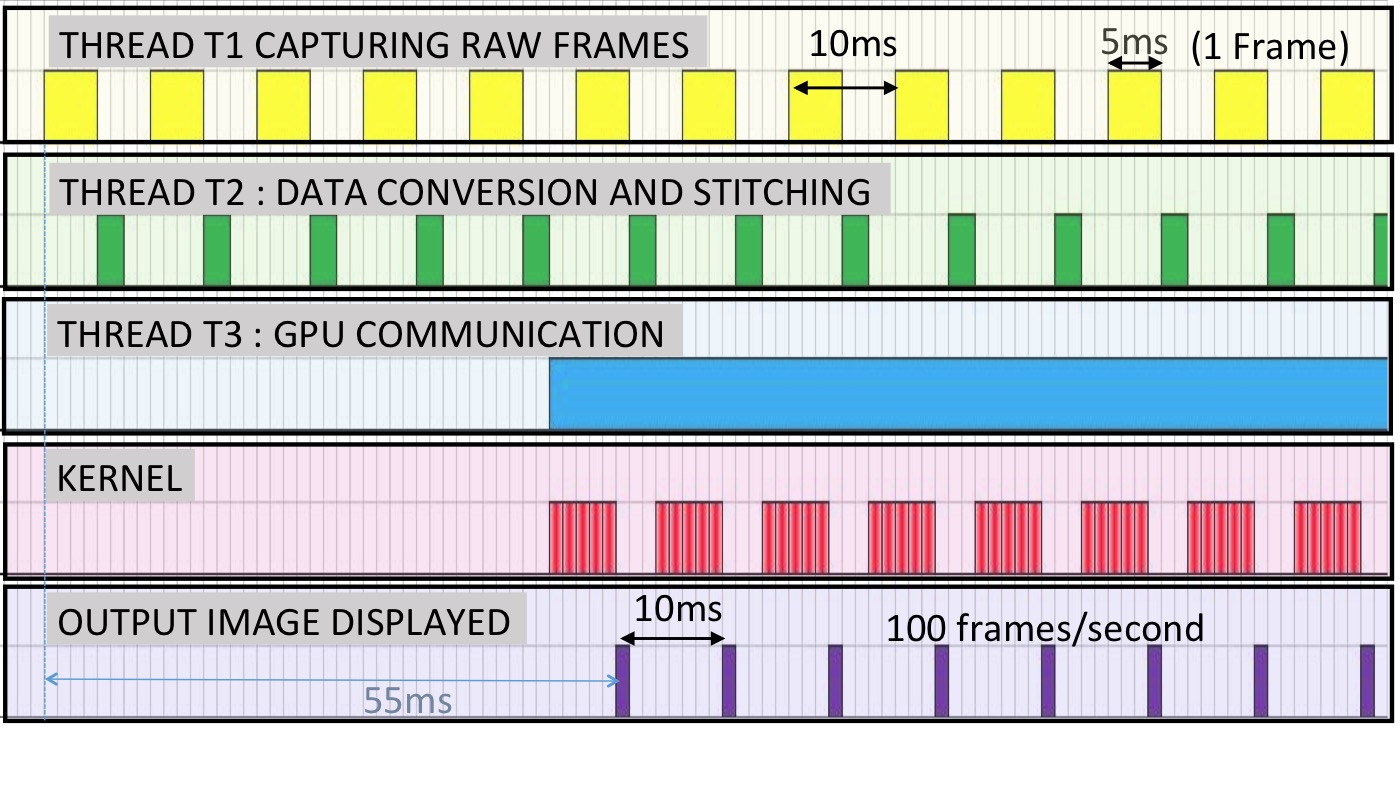}
\caption{ \it {Timing sequence for various constituents of the parallel program. The
times were obtained from an actual run of the experiment using the equipment   described in the text. } In this example frames were acquired over 5ms duration, at 100fps. The processed images were displayed at the same rate.\\ }
\label{fig-timing-sequence}
\end{figure*}

  With multi-threads in the CPU 
simultaneously performing other (non-computational) tasks, it is now 
possible to acquire and to process the 
images simultaneously.  This powerful technique 
allows one to obtain results in ~milliseconds,  faster than the persistence 
of images in the human eye. The timing sequence  is shown in Fig. 
\ref{fig-timing-sequence} for the set of parameters used in the table-top experiments described in the main text. Here frames  of size 600 x 600 pixels 
are acquired 
at  100fps and QLD is performed on data collected over 50ms. The first processed 
image appears in less than 55ms from the start of acquisition of the first frame and 
within 5ms of  acquisition of the last frame.  
Processed images appear  at the same rate as  acquistion. 
Clearly, using  data parallelization in GPU, we can process N frames to obtain one 
single image 
{\it much  before} the next frame (i.e., the $(N + 1)^{th}$ frame) is acquired. 
One would expect the processing would fail to complete when the inter-frame 
interval is small. However, 
the limitation arises due to the camera, rather than the processing - for any given 
frame-size, the camera has a limit on the frame transfer rate; the resulting 
inter-frame interval  is much longer than the time needed for processing. Using a sliding sequence, we display 
frames of size $600\times 600$ pixels at 100fps, thus providing a continuous, flicker-free
display of processed images. This rate reduces as the number of pixels per frame increases; for the full $2560 \times 2160 $ pixels, a processing rate of 20fps is achieved. \\
In addition to the numerous advantages of QLD  mentioned in the main 
text, we point out here several features that makes QLD  
particularly suited for GPU-parallelisation. 
The CUDA FFT library has functions that can only be called 
from host, and thus the program is limited to the number of 
CPU threads, in this case 8.  Writing our 
own kernel for QLD allowed us to execute all of the 
multiplication and accumulation for every pixel at one go. 
In addition, this approach allowed us to free other CPU 
threads for their respective activities rather  than to them 
await the  result, as is the case for   CUDA FFT.
Another distinct advantage of QLD is that it requires a single
buffer that can store one frame ($n^2$ values), while 
FFT, on the other hand,  requires three buffers: an input 
buffer, a process buffer and an output buffer, which scale in size
with the length of the time series. If a sliding window of 
length $N$ is used, the buffer requirement for QLD is 
increased to $Nn^2$, but this is still a factor of 3 smaller 
than  what would be  required for FFT. 

\end{document}